\documentclass[twocolumn,showpacs,preprintnumbers,amsmath,prl]{revtex4}
\usepackage{epsfig}

\begin{document}

\title{Multi-Scaling Comparative Analysis of Time Series \\
and a Discussion on ``Earthquake Conversations" in California}
\author{Nicola Scafetta$^{1}$ and Bruce J. West$^{1,2}$}

\address{$^{1}$ Physics Department, Duke University, Durham, NC 27708} 
\address    {$^{2}$ Mathematics Division,
Army Research Office, Research Triangle Park, NC 27709. }

\begin{abstract}
Time series are characterized by complex memory and/or distribution
patterns. In this letter we show that models obeying to different statistics
may equally reproduce some pattern of a time series. In particular we
discuss the difference between L\'evy-walk and fractal Gaussian intermittent
signals and show that the adoption of complementary scaling analysis
techniques may be useful to distinguish the two cases. Finally, we apply
this methodology to the earthquake occurrences in California and suggest the
possibility that earthquake occurrences are described by a \textit{colored}
(= `long-range correlated') Generalized Poisson model.
\end{abstract}


\date{\today}
\maketitle


Herein we introduce a method of multi-scaling comparative analysis (MSCA)
for the study of intermittent signals. We show that to distinguish between
fractal Gaussian intermittent noise and L\'{e}vy-walk intermittent noise the scaling
results obtained using diffusion entropy analysis (DEA) should be compared
with those obtained from both finite variance scaling methods (FVSM) and
probability distribution functions (pdf) \cite{scafetta1,scalingdetection}.
Finally, we apply MSCA to the seismic data of California and suggest that,
instead of being described by a statistics according to which the waiting
times between Omori's earthquake clusters are uncorrelated from one another,
as the traditional Generalized Poisson model \cite{stain,vito} or a recent L%
\'{e}vy-walk like model assume \cite{vito}, the data may also be
characterized by intercluster \textit{1/f } long-range correlations that may
disclose the \textit{earthquake conversations} recently suggested by Stein 
\cite{stain}.

Hurst, in his pioneering work \cite{Hurst}, introduced the notion of \textit{%
rescaled range analysis} of a time series that takes the scaling form of $%
(R/S)(t)\propto t^{H}$ ($H$ is now called the Hurst exponent). This
stimulated Mandelbrot to introduce the concept of fractional Brownian motion
(fBm) \cite{Mandelbrot}. In a random walk context the value $H=0.5$
indicates uncorrelated noise, $0<H<0.5$ indicates
anti-persistent noise and $0.5<H<1$ indicates
persistent/long-range correlated noise \cite{Mandelbrot}. Alternative
scaling methods applied to a time series $\{\xi _{i}\}$, where $i=1,2,\dots $,
 focus on the autocorrelation function $C(t)\propto \langle \xi _{i}~\xi
_{i+t}\rangle \propto t^{2H-2},$ on the power spectrum representation $%
P_{S}(f)\propto f^{1-2H}~$ \cite{politi} and on the evaluation of the
variance of the diffusion generated by $\{\xi _{i}\}$ \cite{DFA} ( $%
V(t)\propto t^{2H}$). All such scaling methods are related to the original
Hurst's analysis and  yield his H-exponent. These techniques, referred by
us  as FVSM,  assume a finite variance and according to the
central limit theorem (CLT) \cite{gnedenko} the underlying statistics are
Gaussian.

Recently, Scafetta \textit{et al.} \cite{scafetta1} introduced a
complementary scaling analysis, the DEA, that focuses on the scaling
exponent $\delta $ evaluated through the Shannon entropy $S(t)$ of the
diffusion generated by the fluctuations $\{\xi _{i}\}$ of a time series\cite
{scafetta1,scalingdetection}. Here, the pdf of the diffusion process, $p(x,t)
$, is evaluated by means of the subtrajectories $x_{n}(t)=\sum_{i=0}^{t}\xi
_{i+n}$ with $n=0,1,\dots$ The pdf scaling property for a fractal
anomalous diffusion takes the form 
$p(x,t)={t^{-\delta }}~F\left( x~t^{-\delta }\right),$ 
and its entropy increases in time as 
$S(t)=-\int_{-\infty }^{\infty }p(x,t)\ln [p(x,t)]~dx=A+\delta ~\ln (t),$ 
where $A$ is a constant. One can also examine the scaling
properties of the second moment for the same process using the FVSM. One
version of FVSM is the standard deviation analysis (SDA) \cite
{scalingdetection} which is based on the evaluation of the standard
deviation $D(t)$ of the same variable $x$ and pdf $p(x,t)$, and yields 
$D(t)=\sqrt{\langle x^{2};t\rangle -\langle x;t\rangle ^{2}}\propto t^{H}$
\cite{scalingdetection,Mandelbrot}.

Note that the entropy $S(t)$ does not require the variance of the pdf $p(x,t)
$ to be finite \cite{scalingdetection}. The existence of scaling for a
process with a diverging second moment implies that DEA is \textit{%
complementary} to and not simply an \textit{alternative} to FVSM. So, the
scaling exponent $\delta $ is conceptually different from the Hurst exponent 
$H$ measured by means of the FVSM. This suggests that the scaling exponents $%
\delta $ and $H$ may fulfill multiple relations according to the process
under study and, therefore, the combined use of DEA, SDA and pdf analysis
may increase our understanding of complex phenomena through a MSCA.

\begin{figure}[tbp]
\epsfig{file=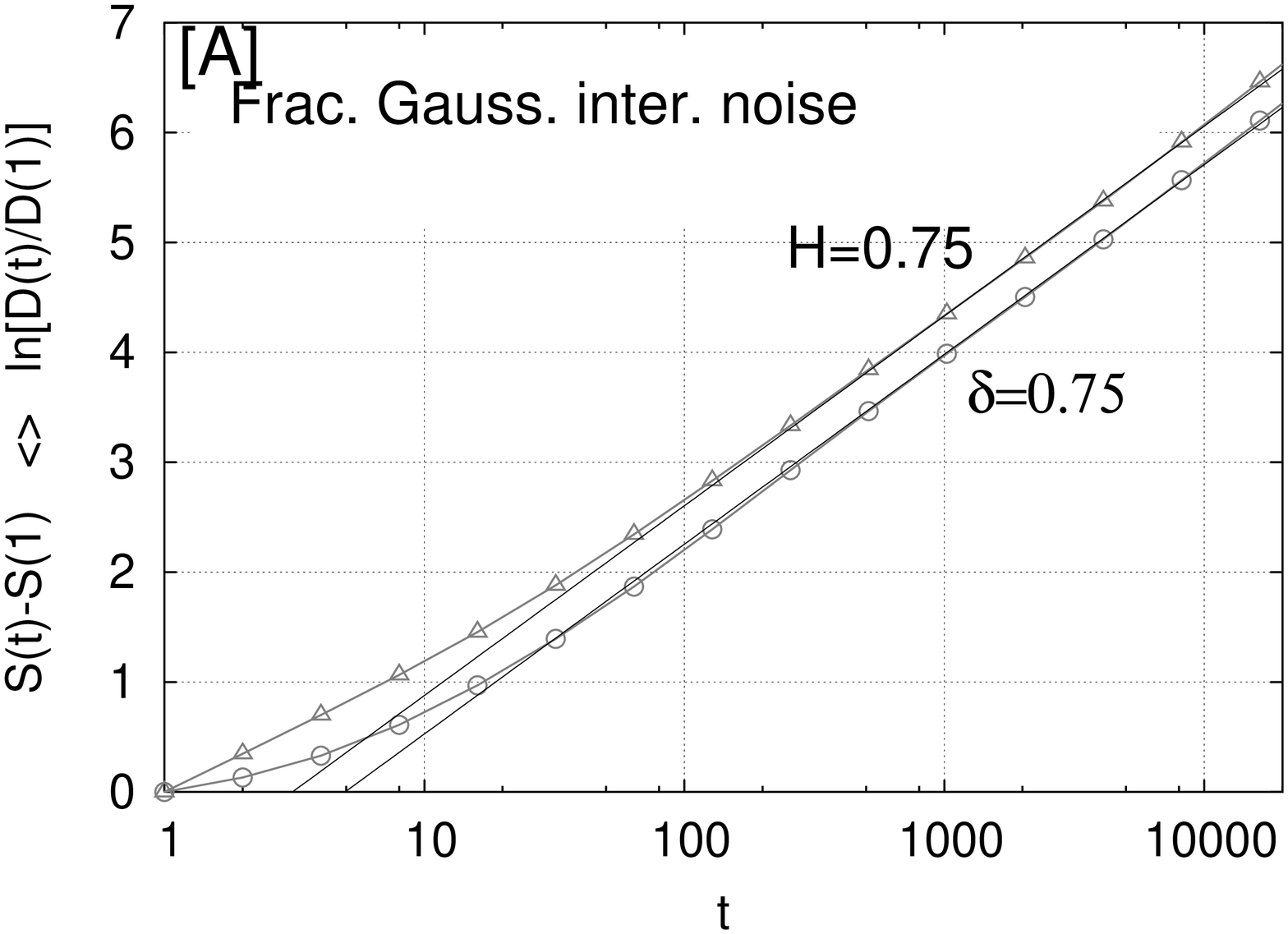,height=4.25cm,width=3.9cm,angle=-90} %
\epsfig{file=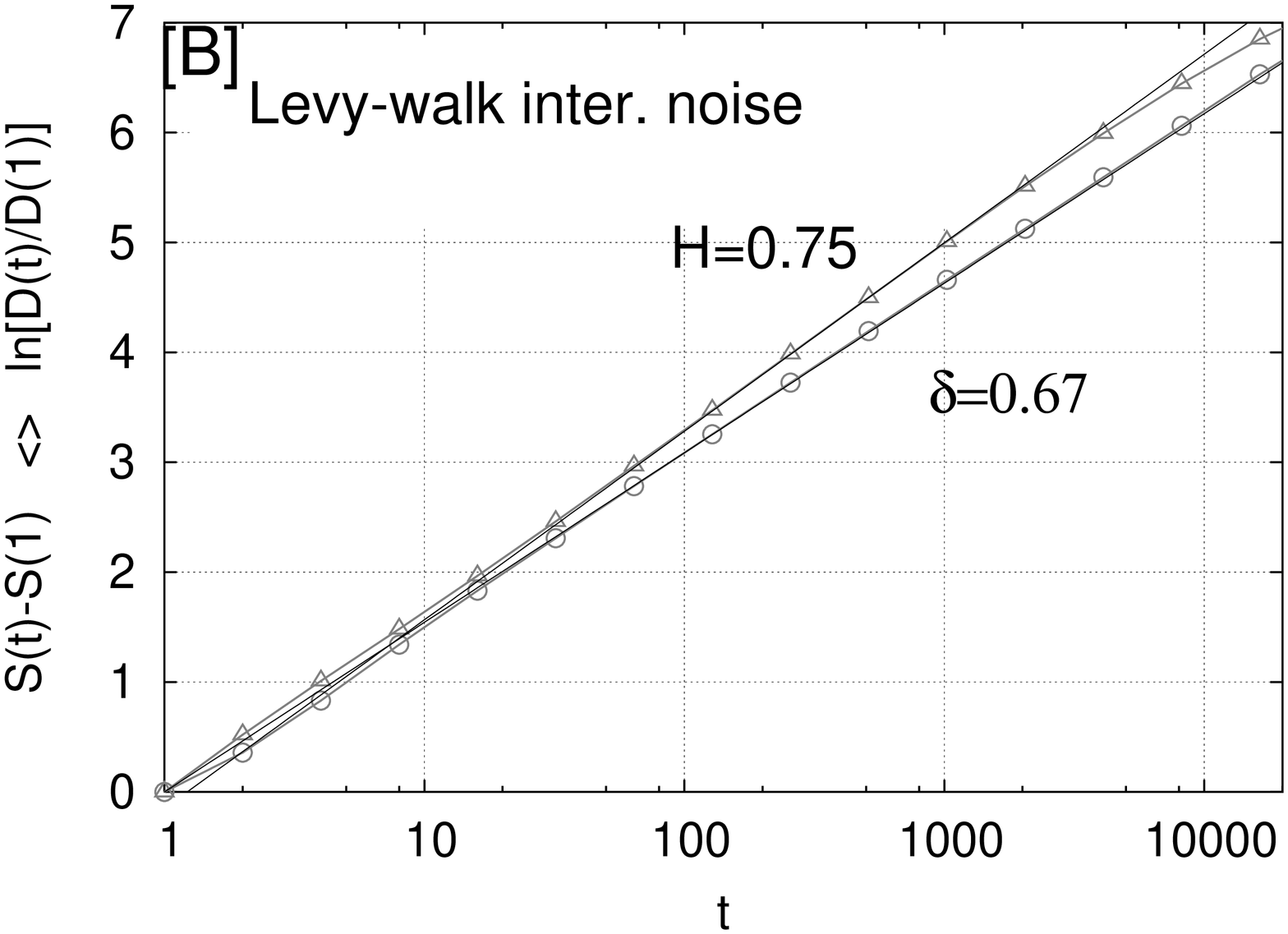,height=4.25cm,width=3.9cm,angle=-90}
\caption{ DEA end SDA of [A] a \textit{fractal Gaussian intermittent noise}
with $\psi(\tau) \propto \exp(- \tau/\gamma)$ with $\gamma=25$ and $%
H=\delta=0.75$; the fractal Gaussian relation (\ref{gaussdh}) of equal
exponents is fulfilled; [B] a \textit{L\'evy-walk intermittent noise} with $%
\psi(\tau)\propto \tau^{-\mu}$ and $\mu=2.5$; note the bifurcation between $%
H=0.75$ and $\delta=0.67$ caused by the L\'evy-walk relation (\ref
{relHdelta34}). }
\end{figure}

Herein we focus on the statistics of intermittent noises. The simplest way
to represent intermittent noise $\{\xi _{i}\}$ is through a dichotomous
representation in which the value $\xi =1$ indicates the occurrence of an
event and the value $\xi =0$ represents no-event \cite
{Palatella,solarflares,dea4,solarflares1}. An intermittent noise is
characterized by the correlation properties of the waiting time sequence $%
\{\tau _{j}\}$ between consecutive events and by its waiting time
distribution $\psi (\tau )$. There are two basic distinct forms of
intermittent noises:\newline
1) \textit{Fractal Gaussian intermittent noise} is characterized by a
long-range correlated waiting time sequence, $\langle \tau _{i}~\tau
_{i+t}\rangle \propto t^{2H-2}$, and by a finite variance waiting time
distribution $\psi (\tau )$ whose form may be, for example, that of a Gaussian,
exponential or Poisson distribution. The diffusion generated by a fractal
Gaussian intermittent noise is a particular type of fBm and satisfies the
asymptotic scaling realtion between indices
\begin{equation}
\delta =H~.  \label{gaussdh}
\end{equation}
We refer to (\ref{gaussdh}) as the \textit{fractal Gaussian diffusion
relation}. If the long-range correlations of $\{\tau _{i}\}$ are destroyed
via shuffling, the new intermittent sequence is characterized by the value $%
H=\delta =0.5$ of random time series. Fig. 1A shows the scaling properties
of a computer generated fractal Gaussian intermittent noise with an
exponential waiting time distribution and $H=\delta =0.75$. \newline
2) \textit{L\'{e}vy-walk intermittent noise} is characterized by an
uncorrelated waiting time sequence, $\langle \tau _{i}~\tau _{j}\rangle
\propto \delta _{ij}$, and a L\'{e}vy or an inverse power law waiting time
distribution 
$\psi (\tau )\propto {(T+\tau )^{-\mu }},$ 
with $2<\mu <3$ that ensures that although the first moment of $\tau $ is
finite, the second moment diverges. The presence of a L\'{e}vy-walk process
in a given time series can be detected \cite{scalingdetection,Palatella} by
mean of the following asymptotic relation among the three exponents 
\begin{equation}
0.5<\delta ={\left( 3-2H\right) }^{-1}={\left( \mu -1\right) }^{-1}<1~.
\label{relHdelta34}
\end{equation}
We refer to (\ref{relHdelta34}) as the \textit{L\'{e}vy-walk diffusion
relation}. Interesting applications of this type of noise have been found in
several L\'{e}vy phenomena including the distribution of solar flares \cite
{Palatella,solarflares,dea4,solarflares1,shlesinger}. In the particular case
in which the sequence $\{\tau _{i}\}$ is correlated the scaling exponents $%
\delta $ and $H$ are larger than the values predicted by Eq. (\ref
{relHdelta34}). Fig. 1B shows the scaling properties of a computer generated
random L\'{e}vy-walk intermittent noise with $\mu =2.5$ that has $H=0.75$
and $\delta =0.67$.

We stress that the L\'{e}vy-walk relation (\ref{relHdelta34}) is fulfilled
if the waiting times $\{\tau _{i}\}$ are uncorrelated, in which case any
shuffling of $\{\tau _{i}\}$ would not alter the scaling exponents $H$ and $%
\delta $. In fact, the super-diffusion scaling exponents $0.5<\delta <H<1$
of a L\'{e}vy-walk intermittent noise are related to the fatness of the waiting time
inverse power law tail, as measured by the exponent $\mu$. Contrary to a
fractal Gaussian intermittent noise, this L\'{e}vy scaling does not imply a
temporal correlation, or a historical memory, among events because the
occurrence of future events is independent of the frequency of past events.

We also observe that there exist particular intermittent sequences obtained
by mixing L\'{e}vy and Gaussian noises \cite{dea4}, with a \textit{L\'{e}vy
memory beyond memory} \cite{palall} or by substituting an event with a
cluster of events \cite{vito}. In these cases the asymptotic properties of
the scaling exponent $H$ and $\delta $ are expected to depend on the
component, Gaussian or L\'{e}vy, with the strongest persistence.

The relations (\ref{gaussdh}) and (\ref{relHdelta34}), and the
correlation/shuffling effects indicate that the DEA should be jointly used 
with the FVSM and/or pdfs. The adoption of
a single technique can lead to a misinterpretation of the characteristics of
a phenomenon, because L\'{e}vy-walk intermittent noise can be confused with
fractal Gaussian intermittent noise, and uncorrelated noise of one kind of
statistics can be mistaken for correlated noise with another kind of
statistics. Figs. 1A and 1B clearly show that the determination
of only one of the two exponents $H$ and $\delta$ is not sufficient to
conclude whether a phenomenon is characterized by a L\'{e}vy-walk
intermittent statistics or by a fractal Gaussian intermittent statistics.
So, we suggest a MSCA by combining complementary techniques.

Recently, DEA has been applied by Mega \textit{et al.} \cite{vito} to study
the time distribution of earthquakes in Southern California ($20^{0}$-$45^{0}
$ N latitude and $100^{0}$-$125^{0}$ W longitude) from 1976 to 2002. The
catalog \cite{web} is complete for local events with magnitude $M\geq 3$
since 1932, for $M\geq 1.8$ since 1981 and for $M\geq 0$ since 1984. The
time intervals between large earthquakes was studied in Ref. \cite{vito} by
setting a temporal variable $\xi (t)=1$ at the occurrence of an earthquake
with a magnitude larger than a given threshold $M_{t}$, and by setting $\xi
(t)=0$ when no earthquake of the specified magnitude occurs. We refer to $%
\{\tau _{i}\}$ as the waiting time sequence between consecutive earthquakes
with $M\geq M_{t}$. So, the intermittent sequence $\xi (t)$ was analyzed by
means of the DEA and the measured scaling exponent was $\delta =0.94\pm 0.01$%
. The authors of Ref. \cite{vito} concluded that the time intervals, $\tau
^{[m]}$, between two consecutive Omori's earthquake clusters \cite{omori} is
modeled by an inverse power law $\psi (\tau ^{[m]})\propto (\tau
^{[m]})^{-\mu }$ with an exponent $\mu =2.06$ calculated via Eq. (\ref
{relHdelta34}). This calculation was based on the traditional assumption 
\cite{vito} that the waiting times between such clusters are uncorrelated ($%
\langle \tau _{i}^{[m]}~\tau _{j}^{[m]}\rangle =\delta _{ij}$, implying that
the observed superdiffusion is induced by a L\'{e}vy-walk between the
Omori's clusters. Finally, Mega \textit{et al.} \cite{vito} showed that a
synthetic sequence produced with Omori's uncorrelated clusters, $%
\langle \tau _{i}^{[m]}~\tau _{j}^{[m]}\rangle =\delta _{ij}$, temporally
distributed according to an inverse power law $\psi (\tau ^{[m]})\propto
(\tau ^{[m]})^{-\mu }$ with $\mu =2.06$, generates a superdiffusive process
with $\delta =0.94$.

However,  the authors of Ref. \cite{vito} did not make the important 
distinction between L\'{e}vy-walk and fractal Gaussian intermittent noises
such as we did above. We showed that a scaling exponent in the range $%
0.5<\delta <1$ can be associated with either a correlated fractal Gaussian
intermittent noise or with an uncorrelated L\'{e}vy-walk noise. Consequently
we apply a MSCA to determine which of the two statistics better describes
the data.

\begin{figure}[tbp]
\epsfig{file=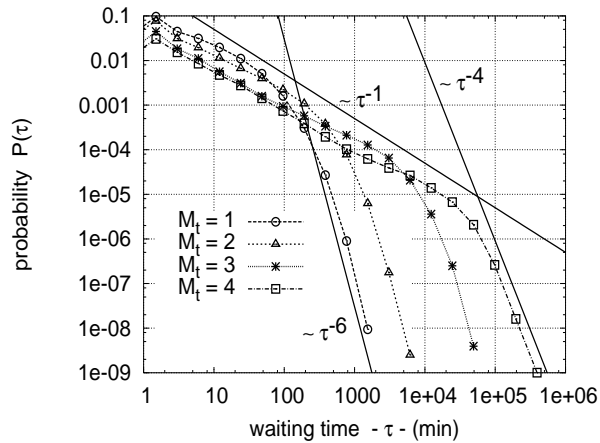,height=8cm,width=6.0cm,angle=-90}
\caption{ pdf of the waiting times $\tau_i$ of earthquakes with a magnitude $%
M\geq M_t$ = 1, 2, 3 and 4. The initial $P(\tau)\propto 1/\tau$ is the
Omori's law \protect\cite{omori}. }
\end{figure}

Fig. 2 shows the waiting time pdfs between earthquakes using four magnitude
thresholds $M_{t} =$ 1, 2, 3 and 4. The pdfs show an initial Omori's law 
\cite{omori} ($P(\tau )\propto 1/\tau $), but the pdf tails  present a large
inverse power law exponent $\mu >4$ and may even approach an exponential (or
Poisson) distribution asymptotically. The Omori's law is determined by the
short-range correlated aftershocks \cite{omori}  and lasts for a time that
increases with the magnitude threshold. If the waiting time distribution
between Omori's clusters were an inverse power law with $\mu=2.06$ it might
be expected that by increasing the magnitude threshold, most of the
aftershocks could be cut off and the tail of the distribution could converge
to an inverse power law with $\mu=2.06$. This does not seem to happen.
Therefore, such an $\mu=2.06$ inverse power law, if it is real, cannot be
observed in this way.

\begin{figure}[tbp]
\epsfig{file=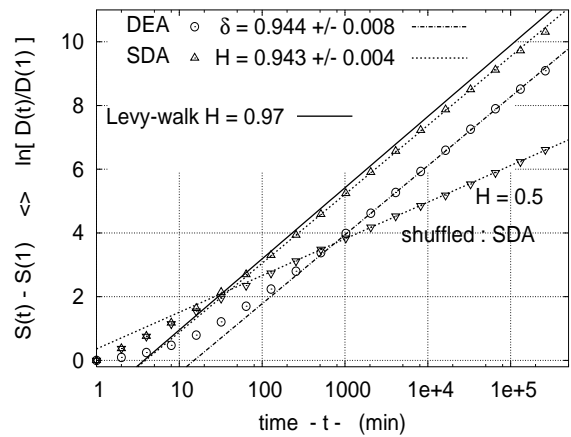,height=8cm,width=6.0cm,angle=-90}
\caption{DEA and SDA of the intermittent time signal $\xi(t)$ for magnitude $%
M\geq M_t$ = 1. The data are fitted with scaling exponents $\delta=0.944 \pm
0.008$ and $H=0.943\pm0.004$. The uppermost solid line with $H=0.97$
corresponds to the expectation of H if the Levy-walk condition (\ref
{relHdelta34}) holds true. }
\end{figure}

Fig. 3 compares the DEA and SDA applied to the intermittent sequence $\xi (t)
$ of earthquakes with $M\geq 1$. Different magnitude thresholds give similar
results. If these data corresponded to a random intermittent L\'{e}vy-walk
and if the curves shown in the figure corresponded to the asymptotic regime,
the condition (\ref{relHdelta34}) interrelating the exponents should hold. A
rigorous DEA fit in the range $\left[ 2^{7}:2^{15}\right] $ gives $\delta
=0.944\pm 0.008$ implying a L\'{e}vy-walk $H=0.97\pm 0.005$. Instead, the
same-range SDA fit gives $H=0.943\pm 0.004.$ The error analysis seems to
confirm better the Gaussian relation of equal scaling exponents (\ref
{gaussdh}) because $\delta $ and $H$ overlap within the statistical error as
in Fig. 1A, while the difference between the measured $H$ and the L\'{e}%
vy-walk $H$ is statistically significant ($p<0.01$). By shuffling
the earthquake waiting time intervals $\{\tau _{i}\}$ we get $H=0.5$.
Finally, by directly applying DEA and SDA to the waiting time series $\{\tau
_{i}\}$, we again get $\delta =H=0.94$. These findings suggest that the data
do not fulfill the L\'{e}vy-walk relation (\ref{relHdelta34}) and that it
might be more likely that the Californian earthquakes are long-range
temporal correlated according to the persistence of a fractal Gaussian
intermittent noise with $H\approx 1$ known as $1/f$ or \textit{pink} noise 
\cite{politi}.

The curve with circles in Fig. 4 shows the DEA applied to a synthetic
earthquake catalog obtained by \textit{coloring} a kind of Generalized
Poisson model for earthquakes. First we generated several Omori's clusters
exactly as done in Ref. \cite{vito}, that is, by assuming that the number of
earthquakes in a cluster follows an inverse power law distribution with
exponent equal to 2.5 and that the events within the same cluster are
temporally distributed according to the Omori's law, that is, an inverse
power law with exponent equal to $p=1$. We generated a total number of
events equal to the total number of earthquakes in the catalog with a
magnitude threshold $M_t=1$. However, contrary to what was done in Ref. \cite
{vito}, we do not randomly ($\langle \tau^{[m]}_i ~ \tau^{[m]}_{j} \rangle
\propto \delta_{ij}$) position these clusters according to an inverse power
law intercluster waiting time distribution $\psi(\tau^{[m]})\propto
1/\tau^{\mu}$ with $\mu=2.06$. Instead,
we distribute the clusters according to a $1/f$ fractal Gaussian
intermittent noise $\{\tau^{[m]}_j\}$ that is, with $\langle \tau^{[m]}_i ~
\tau^{[m]}_{i+t} \rangle \propto t^{2H-2}$ and $H\approx 1$. The
intercluster waiting time distribution $\psi(\tau^{[m]}) \propto
\exp(-\tau^{[m]}/\gamma)$ shown in the insert of Fig. 4 could be substituted
with any other distribution with finite variance. Fig. 4 show that the model
is able to reproduce the same superdiffusion pattern shown by the data.
Finally, the curve with triangles shows the reduction of long-range
persistency of a synthetic catalog obtained with the same clusters of above
but temporally distributed after having shuffled, to randomize, the same
intercluster waiting time sequence $\{\tau^{[m]}_j\}$.

However, Eqs. (\ref{gaussdh}) and (%
\ref{relHdelta34}) are fulfilled only asymptotically where the Central Limit
Theorem for Gaussian processes or its generalization for L\'evy processes
apply \cite{gnedenko}. There might be the possibility that Fig. 3 as well as
figure 2 in Ref. \cite{vito} do not show the asymptotic limit but a
transition regime that is strongly superdiffusive ($\delta \approx 1$)
because of the Omori's intracluster short-range correlations. A long
transition regime is also evident in the curve with triangles shown in Fig.
4 that refers to an uncorrelated intercluster waiting time sequence and
should asymptotically converge to $\delta=0.5$. In fact, also a random
Generalized Poisson model or an ``ETAS" model \cite{Kagan} with appropiate parameters may generate a
synthetic catalog showing superdiffusive properties similar to the real data
within a time range \cite{sornette}. However, we observe that such a result
may depend too strongly on the particular parameters used in the models.

\begin{figure}[tbp]
\epsfig{file=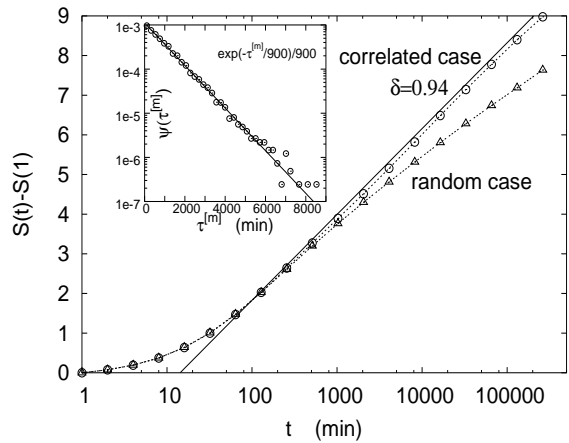,height=8cm,width=6.0cm,angle=-90}
\caption{ DEA applied to a $1/f$ long-range correlated Generalized Poisson
model (circle) for earthquakes for $M\geq 1$. The intercluster waiting time
pdf $\psi(\tau^{[m]})$ is exponential and $\langle \tau^{[m]}_i ~
\tau^{[m]}_{i+t} \rangle \propto t^{2H-2}$ with $H \approx 1$. The curve
with triangles refers to the case in which the intercluster waiting time
sequence $\{\tau^{[m]}\}$ is randomized such that $\langle \tau^{[m]}_i ~
\tau^{[m]}_{j} \rangle \propto \delta_{ij}$.}
\end{figure}

In conclusion, we have discussed some of the difficulties that can be
encountered in interpreting intermittent sequences and shown that models
with alternative statistics can reproduce some pattern of a time series
equally well. This fact suggests the need of an analysis involving
complementary tests. In particular we showed how to distinguish fractal
Gaussian intermittent noise from L\'{e}vy-walk intermittent noise using
MSCA. This methodology has important application in the analysis of
phenomena having intermittent signals because different statistics imply
different dynamics. Our analysis supports the  idea that
earthquakes generate strain diffusion, whose propagation over hundreds of
kilometers induces remote sismic activity \cite{stain,agnes}. This
propagation according to our statistical analysis produces correlations in
the time intervals between earthquake clusters. In fact, the thesis that earthquakes are assembled into uncorrelated Omori's clusters, $\langle \tau _{i}^{[m]}~\tau _{j}^{[m]}\rangle =\delta
_{ij}$, as both the standard \textit{Generalized Poisson} model \cite{vito}
and the L\'{e}vy-walk model \cite{vito} require, seems unrealistic. We
suggest that it is more plausible that earthquake clusters  are $1/f$ long-range correlated and, perhaps, they are subclusters of a larger Omori cluster. In fact, a $1/f$ noise
can be generated by the superposition of  relaxation processes
within a wide range of energies \cite{politi} that may well describe the
coexistent stress alterations caused by old and recent, as well as large and
small shocks. Thus, the $1/f$ long-range correlations may imply that
earthquake occurrences may strongly depend on the geological history of a vast
region. \newline
\textbf{Acknowledgment:} N.S. thanks the ARO for support under grant
DAAG5598D0002.


\end{document}